\documentclass[letterpaper,comsoc]{IEEEtran}

\usepackage{graphicx,epsfig}
\usepackage[noadjust]{cite}
\usepackage{mcite}
\usepackage{amsfonts,helvet}
\usepackage{fancyhdr}
\usepackage{threeparttable}
\usepackage{epsf,epsfig}
\usepackage{amsthm}
\usepackage{amsmath}
\usepackage{siunitx}
\usepackage{amssymb}
\usepackage{booktabs}
\usepackage{dsfont}
\usepackage{subfigure}
\usepackage{color}
\usepackage[linesnumbered,ruled,noend]{algorithm2e}
\usepackage{algpseudocode}
\usepackage{algcompatible}
\usepackage{enumerate}
\usepackage{gensymb}
\usepackage{cancel}
\usepackage{bbm}
\usepackage{graphicx,subfigure}
\usepackage{graphicx}
\usepackage{array}
\usepackage{mathtools}
\usepackage{textcomp}
\usepackage{siunitx}
\newcolumntype{P}[1]{>{\centering\arraybackslash}p{#1}}

\usepackage{eucal}

\makeatletter
\def\blfootnote{\xdef\@thefnmark{}\@footnotetext}
\makeatother

\begin{document}

\title{
    A New Interpretation of the Time-Interleaved ADC Mismatch Problem: A Tracking-Based Hybrid Calibration Approach
}

\author{
    Jiwon Sung
    and Jinseok Choi

    \thanks{
        J. Sung and J. Choi are with the School of Electrical Engineering, Korea Advanced Institute of Science and Technology, Daejeon, South Korea (e-mail: {\texttt{\{jiwonsung, jinseok\}@kaist.ac.kr}}).
    }
}

\maketitle \setcounter{page}{1} 

\begin{abstract}
        Time-interleaved ADCs (TI-ADCs) achieve high sampling rates by interleaving multiple sub-ADCs in parallel. 
        Mismatch errors between the sub-ADCs, however, can significantly degrade the signal quality, which is a main performance bottleneck.
        This paper presents a hybrid calibration approach by interpreting the mismatch problem as a tracking problem, and uses the extended Kalman filter for online estimation and compensation of the mismatch errors.
        After estimation, the desired signal is reconstructed using a truncated fractional delay filter and a high-pass filter.
        Simulations demonstrate that our  algorithm substantially outperforms the existing hybrid calibration method in both mismatch estimation and compensation.
\end{abstract}
\begin{IEEEkeywords}
    TI-ADC, time-varying mismatch errors, hybrid calibration, EKF, fractional delay filter, missing sample problem
\end{IEEEkeywords}
\vspace{-1 em}

\blfootnote{
    © 2025 IEEE. Personal use of this material is permitted. 
    Permission from IEEE must be obtained for all other uses, in any current or 
    future media, including reprinting/republishing this material for advertising 
    or promotional purposes, creating new collective works, for resale or 
    redistribution to servers or lists, or reuse of any copyrighted component 
    of this work in other works.
    
    This work has been accepted for publication in \textit{IEEE Signal Processing Letters} \cite{sung2025tiadc}.
}

\section{Introduction}

Despite the growing demand for high-speed analog-to-digital converters (ADCs) \cite{im2020wireline}, 
the implementation of ADCs that meet such high-speed requirements remains challenging.
An effective solution is to use time-interleaved ADCs (TI-ADCs) \cite{black1980tiadc}, where multiple sub-ADCs are interleaved in parallel.
Here, if a TI-ADC has $M$ sub-ADCs each with a sampling rate of $f_s$, then the TI-ADC will have an overall sampling rate of $Mf_s$.
This approach offers a more practical and cost-effective implementation.
TI-ADCs, however,  introduce their own set of challenges.
The parallel architecture suffers from three main types of mismatch errors between the sub-ADCs: offset mismatch, gain mismatch, and timing mismatch \cite{petraglia1991mismatch, vogel2005mismatch}.
To preserve signal quality, these mismatch errors require precise estimation and subsequent compensation.

Traditional approaches to handling these mismatches fall into two main categories: 
($i$) foreground calibration \cite{schmidt2015foreground, ponnuru2010joint} which interrupts the normal operation of a TI-ADC to insert a known signal as input for estimation,
and ($ii$)  background calibration \cite{tavares2022background, mafi2024digital, razavi2013background} which attempts to estimate mismatch parameters without using a known signal, and instead, utilizes some prior knowledge of the desired signal.
Although foreground calibration is more accurate than using background calibration, the requirement to interrupt the operation of a TI-ADC every once in a while makes it unsuitable for many real-time applications.
On the other hand, background calibration maintains continuous operation but it has a higher computational complexity and requires a lot of samples for estimation, resulting in a much slower convergence rate.

To overcome these limitations, a hybrid calibration approach was proposed in \cite{tsui2013novel}, where a known signal at the receiver is used while not interrupting the operation of a TI-ADC.
In this model, the known signal is sampled instead of the desired signal at every $N_h$ sampling instants.
When a sampling instant is reserved for the known signal, the desired signal is ignored, creating a missing sample.
Consequently, not only do the mismatch errors have to be accurately estimated, but also the signal has to be reconstructed to compensate for both the mismatch errors and the missing samples.
The authors in \cite{tsui2013novel} first expressed the mismatch errors of a TI-ADC with an infinite impulse response (IIR) filter.
Then, the IIR filter was approximated by a practical finite impulse response (FIR) filter that was designed using a variable digital filter (VDF) \cite{farrow1988farrow, pun2002vdf}.
Subsequently, the timing mismatch error can be properly estimated using the normalized least mean squares (NLMS) algorithm \cite{tsui2013novel}.
Although the NLMS algorithm successfully addressed the limitations of the foreground and background calibration techniques by proposing the hybrid calibration approach, its estimation performance is heavily dependent on the computationally-expensive filter optimization of the VDF, and inaccurate approximations of the mismatch effects using the FIR filter can cause imprecise estimation and compensation of mismatch errors.
This limitation motivates us to investigate a method that does not require such an FIR filter optimization.

\begin{figure*}[!t]\centering
	\subfigure{\resizebox{2\columnwidth}{!}{\includegraphics{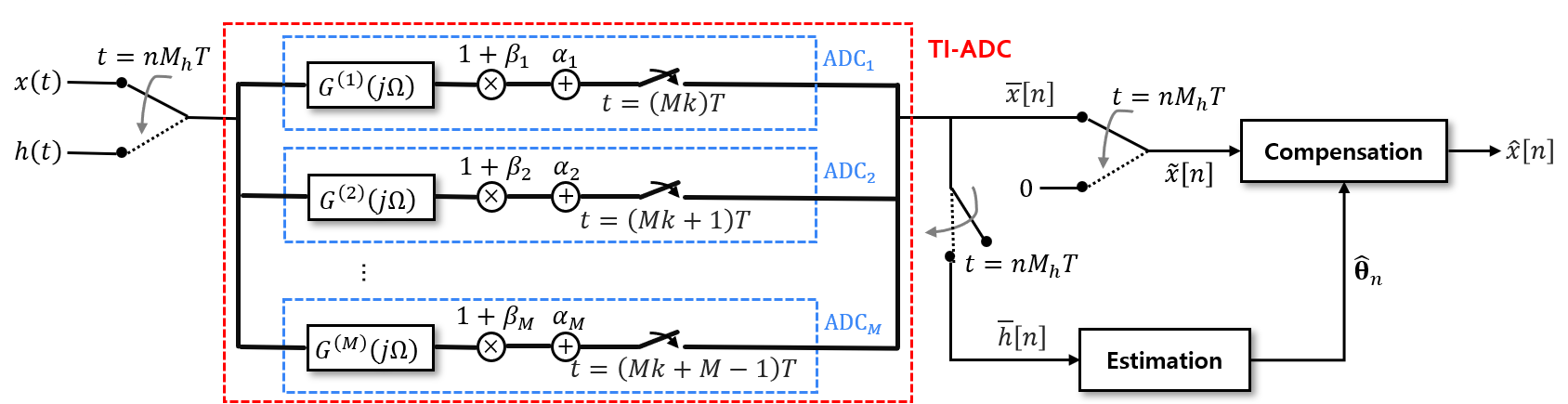}}}
\caption{
    The hybrid calibration structure with mismatch estimation and compensation phases for time-interleaved ADCs.} 
 \label{fig:system_model}
\end{figure*}

In this letter, we interpret the TI-ADC mismatch problem as a tracking problem and propose a hybrid calibration method that nearly achieves the minimum mean square error (MSE) in mismatch estimation.
Via this  interpretation, we  circumvent the need for the optimization of the FIR filter by using the computationally-efficient extended Kalman filter (EKF) \cite{gruber1967ekf, larson1967ekf}, which can be used to accurately track the mismatch errors online.
In addition, unlike \cite{tsui2013novel}, we incorporate the measurement noise for the observation model, which is the TI-ADC output  of the known input signal.
To the best of our knowledge, this is the first study to explicitly consider time-varying mismatch errors as well as the first work to apply a tracking framework.
After we estimate the mismatch errors, we reconstruct the desired signal using a truncated fractional delay filter and a high-pass filter.
The fractional delay filter is used to compensate for the timing mismatch, whereas the high-pass filter is used to compensate for the missing samples.
Simulations demonstrate that our calibration algorithm outperforms the estimation and compensation performance of the NLMS method \cite{tsui2013novel} in the presence of time-varying mismatch errors and measurement noise.

\section{System Model}
\label{sec:sys_model}


We employ a hybrid calibration structure as illustrated in Fig.~\ref{fig:system_model}.
Our hybrid calibration model consists of a TI-ADC, a known signal $h(t)$ generated at the receiver, an estimation module, and a compensation module.
The TI-ADC employs $M$ sub-ADCs that sequentially sample the input signal.
Here, the offset, gain, and timing mismatches of the $m$th sub-ADC are denoted as $\alpha^{(m)}$, $\beta^{(m)}$, and $\phi^{(m)}$, for $m \in \{1, 2, \dots, M\}$.
We note that the timing mismatch $\phi^{(m)}$ represents a fraction of the TI-ADC sampling period $T$, i.e., the timing mismatch in seconds is $\phi^{(m)} T$.
In addition, the timing mismatch of the $m$th sub-ADC can be modeled using a filter, $G^{(m)}(j 2 \pi f)$:
\begin{align}
    \label{eq:freq_response_of_timing_mismatch}
    G^{(m)}(j 2 \pi f) = \mathcal{F} \left\{ \delta \left(t - \phi^{(m)} T \right) \right\} = e^{-j 2 \pi f \phi^{(m)} T},
\end{align}
where $\mathcal{F}$ is the Fourier transform operator and $\delta (\cdot)$ is the Dirac delta function.

The estimation module in Fig.~\ref{fig:system_model} performs online estimation of mismatch errors.
For every $M_h$ samples of the desired signal, the estimation module receives a sample of the known signal $h(t)$ that is generated at the receiver.
More specifically, at time instants $t=n M_h T$, where $n \in \{0, 1, \dots, N-1\}$, the TI-ADC ignores the desired signal $x(t)$ and samples $h(t)$ instead.
Accodingly, it is required for $h(t)$ to be band-limited to $f_s/(2\cdot M\cdot M_h)$ to avoid aliasing.
If we denote $\bar{h}[n]$ as the TI-ADC output of the known signal, it can be modeled as
\begin{align}
    \label{eq:h_bar}
    \bar{h}[n] & = \alpha^{(m')} + \left( 1 + \beta^{(m')} \right) h \left( n M_h T - \phi^{(m')} T \right),
\end{align}
where $m' = \text{mod}(n, M) + 1$.
Here, $\text{mod}(a, b)$ is the modulo operator where $a$ is the dividend and $b$ is the divisor.
Having $m' = \text{mod}(n, M) + 1$ creates a synchronized sampling system where the TI-ADC effectively samples at $M$ times the rate of each individual sub-ADC.
We subsequently compare the TI-ADC output of the known signal $\bar{h}[n]$ to the mismatch-free TI-ADC input of the known signal $h[n] = h(n M_h T)$ to estimate the mismatch errors $\alpha^{(m)}$, $\beta^{(m)}$, and $\phi^{(m)}$.

In addition, the missing samples at time instants $t = nM_hT$ need to be recovered since the desired signal $x(t)$ is ignored at time instants $t=n M_h T$ to obtain the TI-ADC output of the known signal $\bar{h}[n]$ without introducing communication overhead.
For now, we replace the missing samples with zeros:
\begin{align}
    \label{eq:x_tilde}
    \tilde{x}[n] = 
    \begin{cases} 
    0, & n = r M_h, \\
    \bar{x}[n], & \text{otherwise},
    \end{cases}
\end{align}
where $\bar{x}[n]$ is the TI-ADC output of the desired signal with missing samples, $\tilde{x}[n]$ is the output with the missing samples replaced as zeros, and $r$ is an integer.
Fortunately, the missing samples can be reconstructed using a band-limiting assumption \cite{tsui2013novel}.
In this assumption, the desired signal $x(t)$ is oversampled by a ratio of ${\tau} = 2f_{\rm max}/f_s$,
where $f_s = 1/T$, 
and $\tau \in (0, 1)$ and $f_{\text {max}}$ represent the band-limiting parameter and the maximum frequency of the input desired signal $x(t)$, respectively.
By oversampling $x(t)$, we can reconstruct $x(t)$ from $\tilde{x}[n]$ in \eqref{eq:x_tilde} when the missing samples account for up to $(1 - \tau) \times 100\%$ of the total samples according to the sampling theorem \cite{tsui2013novel}.
Hence, to compensate for the missing samples, $\tau$ and $M_h$ must satisfy the following:
\begin{align}
    \label{eq:epsilon_M_h_constraint}
    1 - \tau \geq {1}/{M_h}.
\end{align}

\section{Proposed Algorithm} \label{sec:main}

Now, we are ready to apply a tracking framework to the mismatch estimation problem  and develop a subsequent mismatch estimation and compensation algorithm.

\subsection{State Space Model}

Let ${\pmb{\theta}}_t^{(m)} = [\alpha_t^{(m)}, \beta_t^{(m)}, \phi_t^{(m)}]^\top$ be the state vector representing the mismatch errors of the $m$th sub-ADC at time $t$.
Then, the state space model can be expressed as follows:
\begin{subequations} \label{eq:state_space}
    \begin{align}
        \label{eq:state_evolution}
        {\pmb{\theta}}_t^{(m)} & = u \left({\pmb{\theta}}_{t-1}^{(m)} \right) + \mathbf{e}_t,
        & \mathbf{e}_t \sim \mathcal{N}(\mathbf{0}, \mathbf{Q}), \\
        \label{eq:observation}
        y_t^{(m)} & = \bar{h} \left({\pmb{\theta}}_t^{(m)}, n \right) + v_t, 
        & v_t \sim \mathcal{N}(0, R).
    \end{align}
\end{subequations}
Here, \eqref{eq:state_evolution} and \eqref{eq:observation} are called the state-evolution model and the observation model, respectively.
In the state-evolution model \eqref{eq:state_evolution}, the previous state ${\pmb{\theta}}_{t-1}^{(m)}$ of the $m$th sub-ADC evolves into the state ${\pmb{\theta}}_t^{(m)}$ using a state-evolution function $u(\cdot)$ and an additive white Gaussian noise (AWGN) vector $\mathbf{e}_t \in \mathbb{R}^3$ with covariance matrix $\mathbf{Q}$.
Meanwhile, in the observation model \eqref{eq:observation}, the observation of the $m$th sub-ADC at time $t$ is obtained using the observation function $\bar{h} \left( \cdot \right)$ defined in \eqref{eq:h_bar}, with added AWGN noise $v_t \in \mathbb{R}$ having variance $R$.

In our model, the observation function $\bar{h} \left( \cdot \right)$ is  known.
The state-evolution function $u(\cdot)$, however, may not be exactly known in practice.
In our research, $u(\cdot)$ as well as the noise covariances $\mathbf{Q}$ and $R$ are assumed to be known.
We leave the investigation of a model that operates with unknown state-evolution function and unknown noise statistics as future work.

\subsection{Mismatch Estimation}
\label{subsec:estimation}

In this subsection, we present an EKF-based tracking algorithm for accurate estimation of mismatch errors in TI-ADCs.
There are two main phases that repeat over time: the prediction step and the update step.
At each time step $t$, the algorithm calculates the first and second-order statistical moments of the state ${\pmb{\theta}}_t^{(m)}$ to estimate the state of our system.
From now on, we omit the superscript $(m)$ for brevity.

\subsubsection{Prediction}

In the prediction step, we predict our current estimate of the mean of the state, denoted as ${\hat{\pmb{\theta}}}_{t|t-1}$ and called the \textit{a priori} estimate, based on our previous best estimate ${\hat{\pmb{\theta}}}_{t-1|t-1}$, called the \textit{a posteriori} estimate.
The \textit{a priori} estimate ${\hat{\pmb{\theta}}}_{t|t-1}$ is the prediction of the mean of the state before employing the new measurement data $y_t$ at time $t$,
whereas the \textit{a posteriori} estimate ${\hat{\pmb{\theta}}}_{t|t}$ is the prediction after combining the \textit{a priori} estimate ${\hat{\pmb{\theta}}}_{t|t-1}$  with the new measurement $y_t$.
In order to predict the \textit{a priori} estimate, we obtain the state transition matrix $\mathbf{U}_t$ at time $t$ using the Jacobian of the state-evolution function   $u(\cdot)$ in \eqref{eq:state_evolution} with respect to  ${\hat{\pmb{\theta}}}_{t-1|t-1}$, i.e., $\mathbf{U}_t = \mathcal{J}_{u}\left({\hat{\pmb{\theta}}}_{t-1|t-1}\right)$.
We can then predict the \textit{a priori} estimate of the state, ${\hat{\pmb{\theta}}}_{t|t-1}$ as
\begin{align}
    \label{eq:prediction_state_mean}
    {\hat{\pmb{\theta}}}_{t|t-1} = \mathbf{U}_t {\hat{\pmb{\theta}}}_{t-1|t-1}.
\end{align}
We subsequently calculate how uncertain we are about this prediction by deriving the \textit{a priori} error covariance matrix of the state, denoted as $\mathbf{\Sigma}_{t|t-1}$:
\begin{align}
    \label{eq:prediction_state_covariance}
    \mathbf{\Sigma}_{t|t-1} = \mathbf{U}_t \mathbf{\Sigma}_{t-1|t-1} \mathbf{U}_t^\top + \mathbf{Q}.
\end{align}
Using these \textit{a priori} estimates, we can predict the \textit{a priori} estimate of the observation we expect to see, denoted as $\hat{y}_{t|t-1}$:
\begin{align}
    \label{eq:prediction_observation_mean}
    \hat{y}_{t|t-1} = \bar{h} \left({\hat{\pmb{\theta}}}_{t|t-1}, n\right).
\end{align}
Similar to the \textit{a priori} error covariance matrix $\mathbf{\Sigma}_{t|t-1}$ in \eqref{eq:prediction_state_covariance},
we calculate how uncertain we are about the \textit{a priori} estimate of the observation $\hat{y}_{t|t-1}$ by deriving its \textit{a priori} error covariance $S_{t|t-1}$:
\begin{align}
    \label{eq:prediction_observation_covariance}
    S_{t|t-1} = \mathbf{H}_t \mathbf{\Sigma}_{t|t-1} \mathbf{H}_t^\top + R,
\end{align}
where ${\bf H}_t$ is the Jacobian of the observation function $\bar{h}(\cdot)$ in \eqref{eq:observation} with respect to ${\hat{\pmb{\theta}}}_{t|t-1}$, i.e., $\mathbf{H}_t = \mathcal{J}_{\bar{h}}\left({\hat{\pmb{\theta}}}_{t|t-1}, n\right)$.

\subsubsection{Update}

In the update step, we use the observation $y_t$ at time $t$ as well as the \textit{a priori} estimates ${\hat{\pmb{\theta}}}_{t|t-1}$, $\mathbf{\Sigma}_{t|t-1}$, $\hat{y}_{t|t-1}$, and $S_{t|t-1}$ that we obtained in the prediction step to predict the \textit{a posteriori} estimates of the mean of the state ${\hat{\pmb{\theta}}}_{t|t}$ and the error covariance matrix $\mathbf{\Sigma}_{t|t}$.
We obtain ${\hat{\pmb{\theta}}}_{t|t}$ using the difference between what we observed and what we predicted without the observation:
\begin{align}
    \label{eq:update_state_mean}
    {\hat{\pmb{\theta}}}_{t|t} = {\hat{\pmb{\theta}}}_{t|t-1} + \mathbf{K}_t \Delta y_t,
\end{align}
where the Kalman gain $\mathbf{K}_t$  is  given as
\begin{align}
    \label{eq:kalman_gain}
    \mathbf{K}_t = \mathbf{\Sigma}_{t|t-1} \mathbf{H}_t S_{t|t-1}^{-1},
\end{align}
and $\Delta y_t =y_t - \hat{y}_{t|t-1}$ is the innovation at time $t$.
Lastly, the \textit{a posteriori} estimate of the error covariance matrix $\mathbf{\Sigma}_{t|t}$ can be derived as
\begin{align}
    \label{eq:update_state_covariance}
    \mathbf{\Sigma}_{t|t} = \mathbf{\Sigma}_{t|t-1} - \mathbf{K}_t S_{t|t-1} \mathbf{K}_t^\top.
\end{align}
The EKF algorithm repeats the prediction step and the update step for each new observation $y_t$ in the system to estimate ${\pmb{\theta}}_t$.

\subsection{Mismatch and Missing Sample Compensation}

Given the mismatch estimates from Section~\ref{subsec:estimation}, we now perform signal compensation:  mismatch error correction and  missing sample reconstruction.

\subsubsection{Offset and gain mismatch compensation}

We first compensate for the offset and gain mismatch errors.
Given that the estimation of these mismatch errors are accurate, they can easily be compensated from $\tilde{x}[n]$ in \eqref{eq:x_tilde} as follows:
\begin{align}
    \label{eq:x_tilde_comp}
    \tilde{x}_{\text{comp}}[n] = 
    \begin{cases} 
    0, & n = r M_h, \\
    \left( \bar{x}[n] - \hat{\alpha}^{(m')} \right) / \left( 1 + \hat{\beta}^{(m')} \right), & \text{otherwise},
    \end{cases}
\end{align}
where $\tilde{x}_{\text{comp}}[n]$ is the signal in \eqref{eq:x_tilde} after compensating for the offset and gain mismatch errors, and $m' = \text{mod}(n, M) + 1$.

\subsubsection{Timing mismatch and missing sample compensation}

To compensate for the timing mismatch errors, we design FIR filters that model the timing mismatches using fractional delay filters. 
Let $g_n[k]$ denote the impulse response of a fractional delay filter of length $2 N_g + 1$.
Since the inverse discrete-time Fourier Transform (DTFT) of the frequency response of the timing mismatch of the $m$th sub-ADC in \eqref{eq:freq_response_of_timing_mismatch} is $\text{sinc} \left( n - \phi^{(m)} \right)$, we design the truncated filter for $k = -N_g, ..., N_g$ as
\begin{align}
    \label{eq:truncated_FDF}
    g_n[k] = \text{sinc} \left( k - \phi^{ \left( m' \right) } \right), \quad m' = \text{mod}(n, M) + 1.
\end{align}
For $n \neq r M_h$, we can use this filter to mimic the timing mismatch errors present in \eqref{eq:x_tilde_comp}: 
\begin{align}
    \label{eq:x_comp_desired}
    \tilde{x}_{\text{comp}}[n] \approx \sum_{k=-N_g}^{N_g} g_n[k] \cdot x[n-k], \quad n \neq r M_h
\end{align}

Now, we follow similar compensation steps as in \cite{tsui2013novel}, but using a different FIR filter $g_n[k]$ in \eqref{eq:x_comp_desired}.
Recall from Section~\ref{sec:sys_model} that the missing samples can be compensated if \eqref{eq:epsilon_M_h_constraint} is satisfied.
As the input desired signal $x(t)$ is oversampled by a ratio of $1/\tau$, the DTFT of $x[n] = x(nT)$ is zero for $\tau \pi \leq |2 \pi f T| \leq \pi$.
We utilize this additional information to compensate for the missing samples by using a high-pass filter.
Let $w[k]$ denote the impulse response of the high-pass filter of length $2 N_w + 1$ and a cutoff frequency of $2 \pi f T = \tau \pi$.
Then, $\sum_{k=-N_w}^{N_w} w[k] \cdot x[n-k] \approx 0$.
The filter is implemented as a type I linear-phase FIR filter, which has symmetric coefficients and an odd-numbered filter length.
Since \eqref{eq:x_tilde_comp} gives us $\tilde{x}_{\text{comp}}[n] = 0$ for $n = r M_h$, we apply this filter to $x[n]$ to make use of the additional information that its DTFT is zero for $\tau \pi \leq |2 \pi f T| \leq \pi$: 
\begin{align}
    \label{eq:x_comp_missing_sample}
    \tilde{x}_{\text{comp}}[n] \approx \sum_{k=-N_w}^{N_w} w[k] \cdot x[n-k], \quad n = r M_h.
\end{align}

Next, we present a reconstruction algorithm that compensates for the timing mismatch and the missing samples using a combination of our truncated fractional delay filter $g_n[k]$ and our high-pass filter $w[k]$:
\begin{align}
    \label{eq:g_tilde}
    \tilde{g}_n[k] = 
    \begin{cases} 
    w[k], & n = r M_h, \\
    g_n[k], & \text{otherwise},
    \end{cases}
\end{align}
where $\tilde{g}_n[k]$ is the combined filter.
Although $\tilde{g}_n[k]$ is presented as a non-causal filter, we note that it can easily be implemented as a causal system in practice using sufficient delays.
Using $\tilde{g}_n[k]$, \eqref{eq:x_comp_desired} and \eqref{eq:x_comp_missing_sample} can be combined as
\begin{align}
    \tilde{x}_{\text{comp}}[n]
    \approx \sum_k \tilde{g}_n[k] \cdot x[n-k],
    \quad \forall n.
\end{align} 
Here, our goal is to fully retrieve $x[n]$ from $\tilde{x}_{\text{comp}}[n]$.
However, this typically requires a computationally-expensive matrix inversion.
To design a low-complexity reconstruction algorithm by avoiding a direct matrix inversion, we use an iterative approach called the Gauss-Seidel iteration (GSI) algorithm \cite{tsui2011gsi,tsui2013novel}, which delivers superior convergence rates compared to other iterative approaches.
The GSI algorithm for $n = 0, 1, \dots, N-1$ is as follows:
\begin{align}
    \label{eq:gsi}
    \hat{x}^{(i+1)}[n]
    \nonumber
    & = \tilde{g}_n^{-1}[0] \Bigg( \tilde{x}_{\text{comp}}[n] - \sum_{k=1}^{N_g}  \tilde{g}_n[k] \cdot \hat{x}^{(i+1)}[n-k] \\
    & \quad \quad \quad \quad \quad - \sum_{k=-N_g}^{-1}  \tilde{g}_n[k] \cdot \hat{x}^{(i)}[n-k] \Bigg),
\end{align}
where $\hat{x}^{(i)}[n]$ is the reconstructed signal after the $i$th iteration.
Here, $\hat{x}^{(0)}[n]$ is initialized as $\tilde{x}_{\text{comp}}[n]$.

We remark that unlike \cite{tsui2013novel}, the mismatch estimation can be performed independently from the truncated fractional delay filter, and hence does not require the optimization of the filter for mismatch estimation.
The VDF approach in \cite{tsui2013novel}, however, depends heavily on the performance of a minimax filter optimization, and it directly affects the performance of the mismatch estimation.
In this regard, using our tracking-based approach can be more robust and practical.

\section{Numerical Results}

In our simulations, we evaluate the proposed algorithm by comparing with the existing hybrid calibration approach, the NLMS algorithm, in \cite{tsui2013novel}.
The known signal $h(t)$ is generated as $h(t) = \cos \left( 2\pi f_h t \right)$, where $f_h$ is the frequency and the normalized angular frequency is set as $2 \pi f_h T = 0.8 \pi / (MM_h)$.
Using \eqref{eq:h_bar}, we derive $\mathbf{H}_t$ in \eqref{eq:prediction_observation_covariance} and \eqref{eq:kalman_gain} as
\begin{align}
    \label{eq:observation_jacobian}
    \mathbf{H}_t 
    & = \Big[
    1, \ 
    \cos\left( 2\pi f_h M_h T n - 2\pi f_h T \hat{\phi}_{t|t-1} \right), \\
    \nonumber
    & \ \ \quad (2\pi f_h T) \! \left( 1 \! + \! \hat{\beta}_{t|t-1} \right) 
    \sin \! \left( 2\pi f_h M_h T n - 2\pi f_h T \hat{\phi}_{t|t-1} \right) \!
    \Big].
\end{align}
The rest of the system model parameters are specified as follows:  
the number of sub-ADCs $M$ is $4$, the period of the known signal $M_h$ is $17$, 
the band-limiting parameter $\tau$ is $0.8$,
and the parameters for the compensation filters are $N_g = 20$ and $N_w = 51$.
The observation noise variance is set as $R = 5 \times 10^{-5}$ so that the observation signal-to-noise ratio (SNR) is $50$ dB.
The initial values of the mismatch errors are set as $\alpha_k = [-0.03, 0.05, -0.08, -0.02]$, $\beta_k = [0.05, -0.04, 0.02, -0.09]$, and $\phi_k = [-0.01, -0.05, 0.04, -0.03]$,
and their initial estimates are set as zero for all $M$ sub-ADCs.
For reconstruction, we performed $4$ iterations of the GSI algorithm.

\begin{figure}[!t]\centering
    \begin{subfigure}
        [Estimation NMSE]{\resizebox{0.8\columnwidth}{!}{\includegraphics{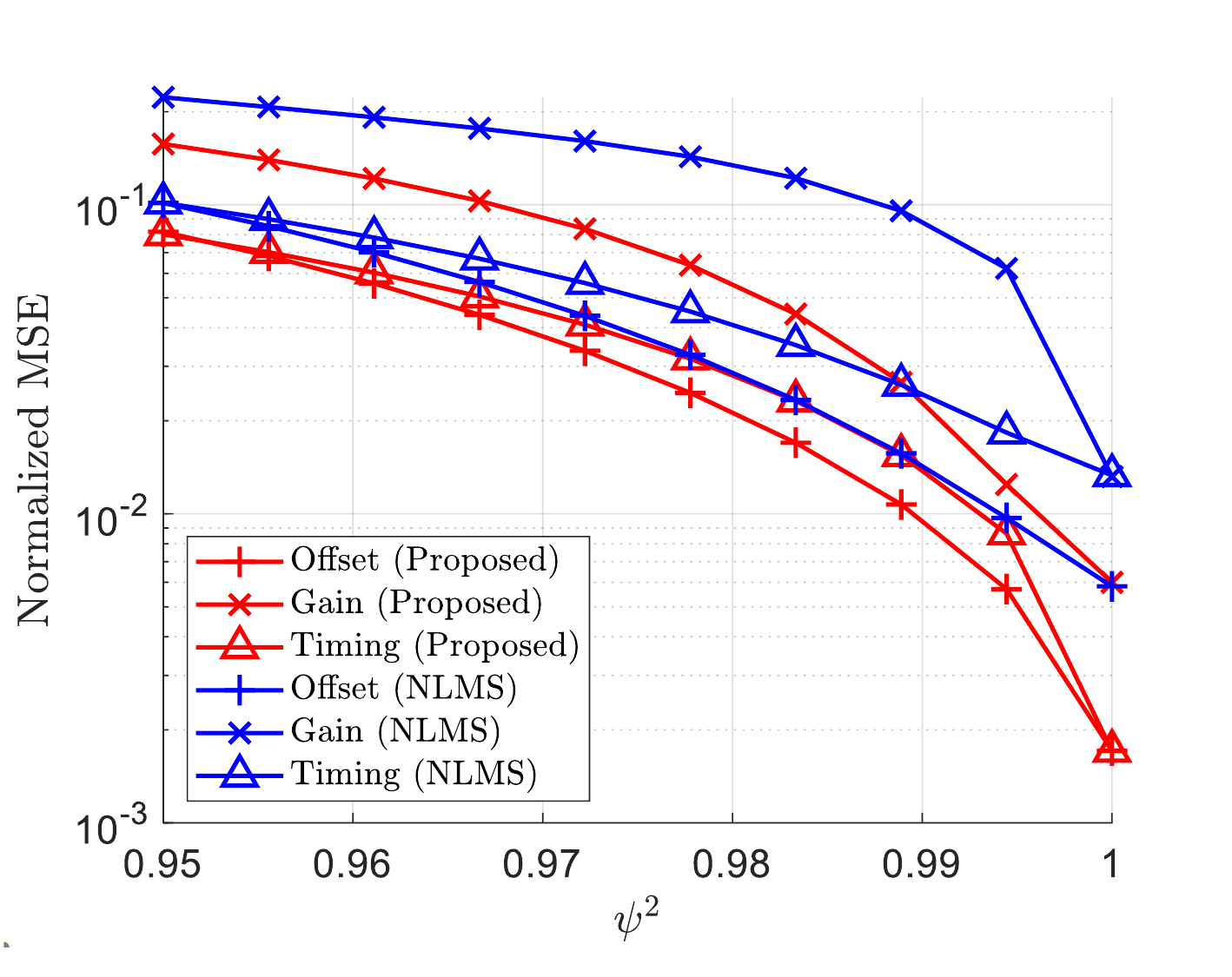}}}
    \end{subfigure}
    \begin{subfigure}
        [Reconstruction NMSE]{\resizebox{0.8\columnwidth}{!}{\includegraphics{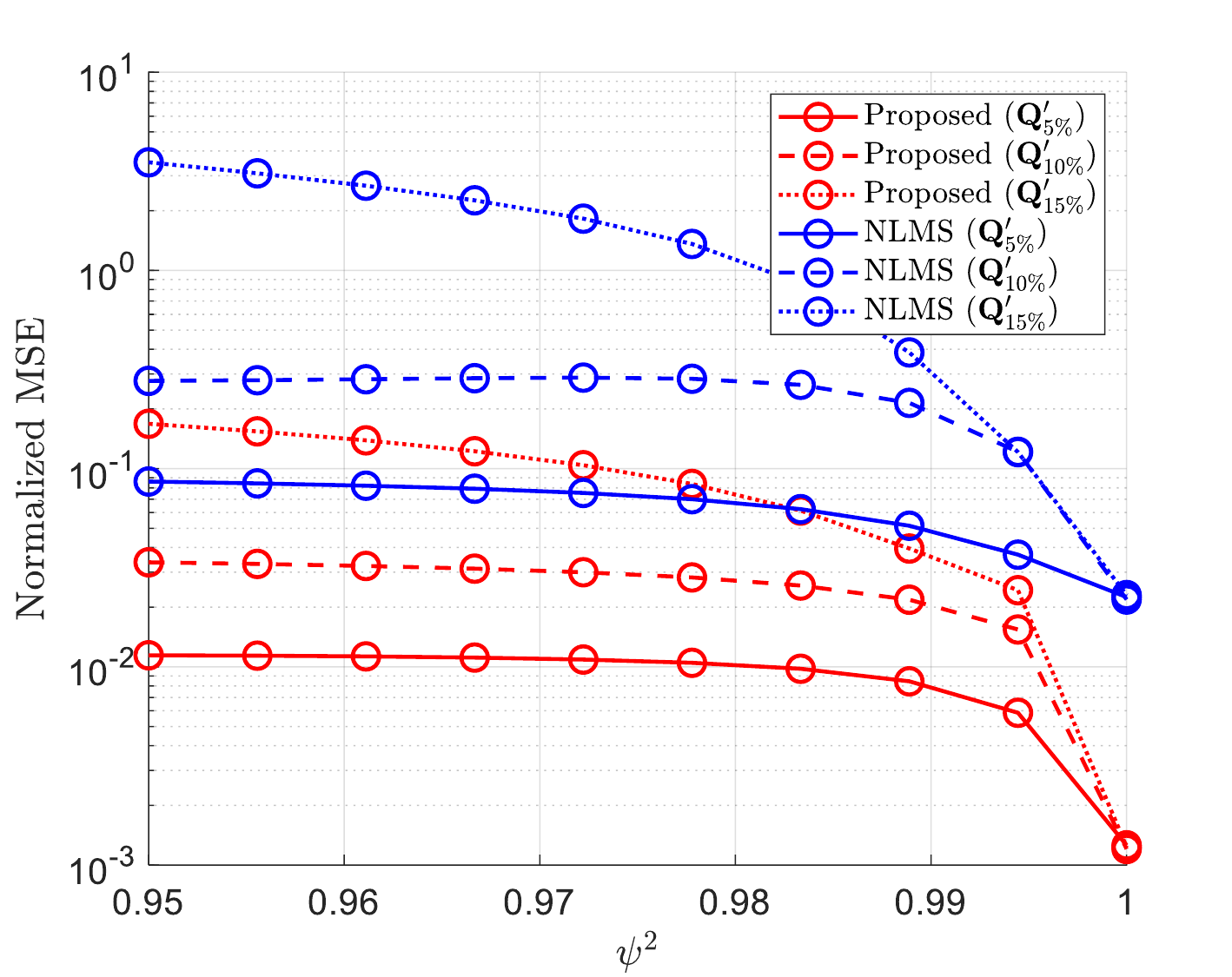}}}
    \end{subfigure}
\caption{
    The normalized MSE of (a) the estimated mismatch errors for $\mathbf{Q}' = \mathbf{Q}'_{5\%}$ and (b) the reconstructed signal for various $\mathbf{Q}'$, both with respect to $\psi^2$.}
\label{fig:NMSE}
\end{figure}

We characterize the time-varying mismatch errors described in \eqref{eq:state_evolution} using an AR(1) model: ${\pmb{\theta}}_t = \mathbf{\Psi} {\pmb{\theta}}_{t-1} + \mathbf{e}_t$.
We note that AR models are commonly used in stochastic processes \cite{haykin1996ar1} and that the transition matrix $\mathbf{\Psi}$ can be easily obtained using training samples in practice.
For the sake of presentation, we assume that the offset, gain, and timing mismatches are uncorrelated with each other and simplify the AR(1) model as
\begin{align}
    \label{eq:AR1}
    {\pmb{\theta}}_t = \psi {\pmb{\theta}}_{t-1} + \sqrt{1-\psi^2} \mathbf{e}_t', \quad \quad \mathbf{e}_t' \sim \mathcal{N}(\mathbf{0}, \mathbf{Q}'),
\end{align}
where $\psi \in [0, 1]$ is the autoregressive coefficient,
and $\mathbf{Q}$ in \eqref{eq:state_evolution} is $\mathbf{Q} = \left(1 - \psi^2 \right) \mathbf{Q}'$.
We evaluate the algorithm for $\mathbf{Q}' \in \{\mathbf{Q}'_{5\%}, \mathbf{Q}'_{10\%}, \mathbf{Q}'_{15\%}\}$,
where $\mathbf{Q}'_{5\%}$, $\mathbf{Q}'_{10\%}$, and $\mathbf{Q}'_{15\%}$ are the amount of the variance of  a uniform random variable with the intervals $[-0.05, 0.05]$, $[-0.1, 0.1]$, and $[-0.15, 0.15]$, respectively, so that the variance of ${\pmb \theta}_t$ is equivalent to those uniform random cases for $\psi \neq 1$.
In addition, as the state-evolution function in \eqref{eq:state_evolution} is $u \left({\pmb{\theta}}_{t-1} \right) = \psi {\pmb{\theta}}_{t-1}$, the state transition matrix $\mathbf{U}_t$ in \eqref{eq:prediction_state_mean} and \eqref{eq:prediction_state_covariance} can be expressed as $\mathbf{U}_t = \mathcal{J}_{u}\left({\hat{\pmb{\theta}}}_{t-1|t-1}\right) = \psi \mathbf{I}_{3}.$
We remark that $\psi$ is typically close to one due to the quasi-static nature of mismatch errors.
Although the case of $\psi=1$ is used by most, if not all, existing baselines in their estimation of mismatch errors, our algorithm works with arbitrary $\psi$.

Fig.~\ref{fig:NMSE}(a) illustrates the normalized MSE (NMSE) of the estimated mismatch errors with respect to $\psi^2$ for $\mathbf{Q}' = \mathbf{Q}'_{5\%}$. 
The number of samples is $N=10^4$ and the desired signal is generated as $x(t) = \sum_{i=1}^{10} \cos(2 \pi f_i t)$, where $2 \pi f_i T = 2i \pi / 25$.
Our results in Fig.~\ref{fig:NMSE}(a) show that our estimation module demonstrates better estimation performance for all mismatch errors.
Additionally, inaccurate estimation of the timing mismatch can substantially affect the gain mismatch estimation, as demonstrated by the relationship in \eqref{eq:h_bar}.
Meanwhile, Fig.~\ref{fig:NMSE}(b) depicts the NMSE of the reconstructed desired signal with respect to $\psi^2$ for various $\mathbf{Q}'$.
As shown in Fig.~\ref{fig:NMSE}(b), our algorithm achieves approximately $10\times$  lower reconstruction error thanks to the enhanced estimation accuracy and the use of a truncated fractional delay filter.
Unlike the baseline approach, our filter design eliminates the need for optimization procedures, further improving efficiency.
Lastly, the numerical results presented in Fig.~\ref{fig:NMSE} demonstrate our algorithm's robust performance across diverse time-varying mismatch errors and measurement noise compared to the benchmark.

\begin{figure}[!t]\centering
	\subfigure{\resizebox{0.8\columnwidth}{!}{\includegraphics{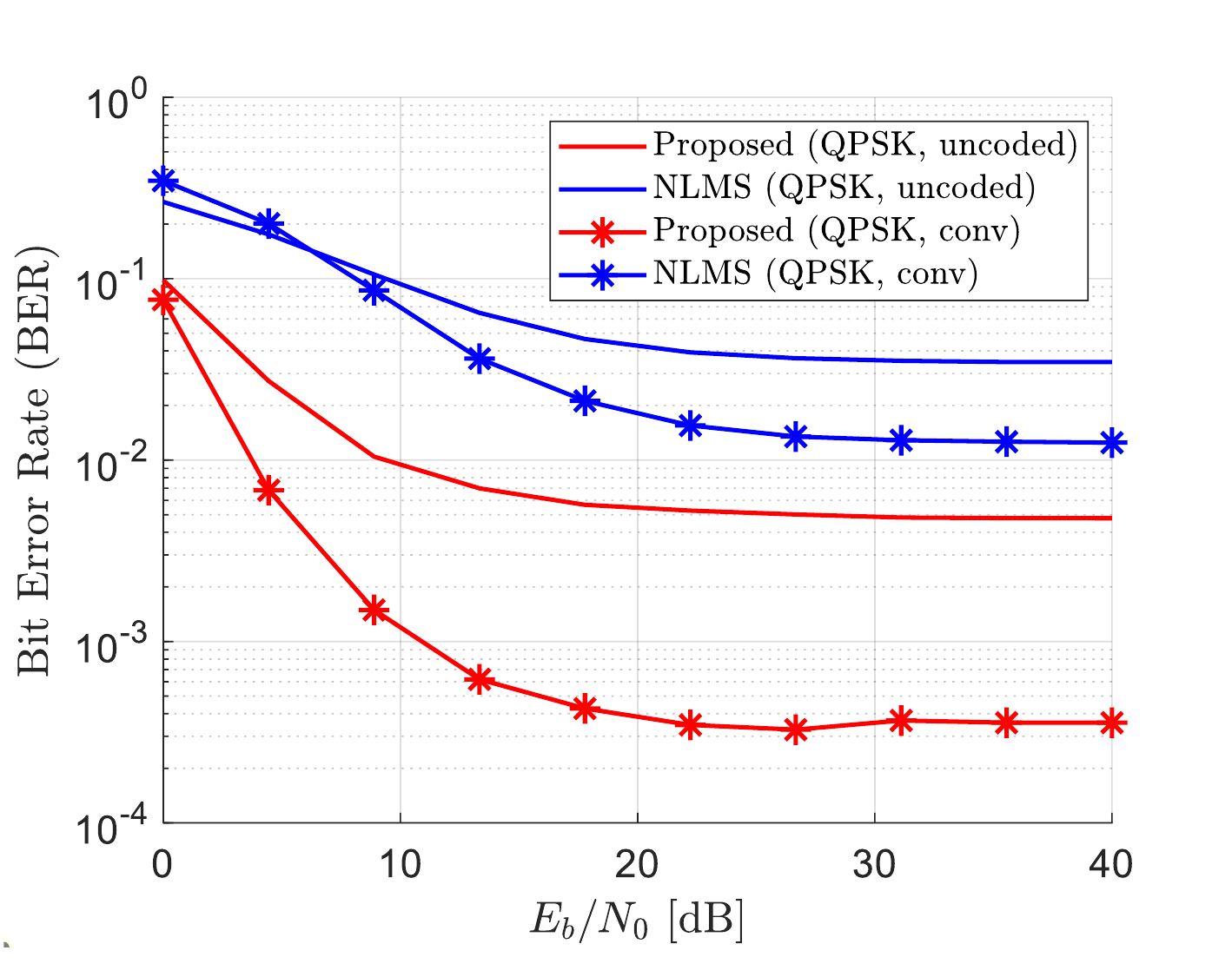}}}
\caption{
    The bit error rate (BER) with respect to the $E_b/N_0$ for $\mathbf{Q}' = \mathbf{Q}'_{5\%}$.} 
 \label{fig:EbNo_vs_BER}
\end{figure}

In Fig.~\ref{fig:EbNo_vs_BER}, we plot the bit error rate (BER) with respect to the energy per bit to noise power spectral density ratio ($E_b/N_0$) in dB scale for $\mathbf{Q}' = \mathbf{Q}'_{5\%}$ and $\psi^2 = 0.99$.
In this simulation, we use Quadrature Phase-Shift Keying (QPSK) for modulation, and transmit the interpolated signal through an AWGN channel.
Then, the noisy signal is sampled by two TI-ADCs: one for the in-phase component and another for the quadrature component.
We evaluate the BER performance under two scenarios: an uncoded system and a system employing rate-$1/2$ convolutional coding with generator polynomials [6,7] in octal notation.
For the coded system, we use a block length of 1000 bits and implement a hard-decision Viterbi decoder at the receiver.
As shown in Fig.~\ref{fig:EbNo_vs_BER}, our proposed algorithm exhibits superior BER performance across both the uncoded and the coded systems.
An interesting observation here is that both our algorithm and the baseline exhibit a BER performance floor, where further increasing $E_b/N_0$ does not yield additional improvements in the BER.
This behavior is attributable to the residual mismatch errors in TI-ADCs after compensation.
Since these mismatch errors cannot be perfectly compensated, the resulting reconstruction error emerges as the dominant source of error when the AWGN channel noise is negligible.
We remark that we used a very simple trellis design for the convolutional code in this simulation, and that further BER improvements could be achieved through the adoption of more sophisticated coding schemes.

\section{Conclusion}

In this work, we have developed a hybrid calibration algorithm for correcting mismatch errors in TI-ADCs using an EKF-based tracking method and a combination of a truncated fractional delay filter and a high-pass filter.
By formulating the estimation problem within a tracking problem framework, we have demonstrated that our EKF-based method achieves superior estimation accuracy compared to the existing hybrid calibration method, and also maintains higher performance with time-varying mismatch errors and measurement noise - a scenario not previously addressed in the literature.
Furthermore, our estimation and compensation modules do not require complex filter optimizations, resulting in a more practical and computationally efficient implementation.
These advantages make our calibration algorithm a promising solution for next-generation high-speed ADC systems where precise calibration is critical for maintaining signal quality.

\bibliographystyle{IEEEtran}
\bibliography{Ref}

\end{document}